\documentclass[12pt]{aastex}
\usepackage{mymanual,onecolfloat,graphicx,enumerate}

\def\etc{{\it etc.}}

\def\pmb#1{\setbox0=\hbox{$#1$}%
  \kern-0.25em\copy0\kern-\wd0
  \kern.05em\copy0\kern-\wd0
  \kern-0.025em\raise.0433em\box0}
\def\spmb#1{\setbox1=\hbox{${\scriptstyle #1}$}%
  \kern-0.25em\copy1\kern-\wd1
  \kern.05em\copy1\kern-\wd1
  \kern-0.025em\raise.0433em\box1}

\def\routine#1{{\tt #1}}
\def\underscore{\underline{\enspace}}

\def\boxit#1{\hbox to 1.5cm{\hfil $#1$}}

\submitted{{}}

\begin{document}

\title{{\tt GALAXY} package for $N$-body simulation}
\author{J. A. Sellwood}
\smallskip
\affil{Dept.\ of Physics \& Astronomy, Rutgers University, \\
136 Frelinghuysen Road, Piscataway, NJ 08854, USA}

\parskip=6pt
\bigskip

\begin{abstract}
This posting announces public availability of the \routine{GALAXY}
software package developed by the author over the past 40 years.  It
is a highly efficient code for the evolution of (almost) isolated,
collisionless stellar systems, both disk-like and ellipsoidal.  In
addition to the $N$-body code {\tt galaxy}, which offers eleven
different methods to compute the gravitational accelerations, the
package also includes sophisticated set-up and analysis software.
This paper gives an outline of the contents of the package and
provides links to the source code and a comprehensive on-line manual.
While not as versatile as tree codes, the particle-mesh methods in
this package are shown, for certain restricted applications, to be
between 50 and 200 times faster than a widely-used tree code.
\end{abstract}

\section{Introduction}
\label{sec.intro}
This publication announces the \routine{GALAXY} software
package\footnote{\copyright\ 2014.  The copyright for this package is
  owned by the author} written almost exclusively by the author
for collisionless simulations of galaxies.  I have developed the code
over a period of almost 4 decades, and continue to make refinements;
it has now reached version 14.  Pieces of the code have been described
in a number of papers in the literature over the years
\citep[][\etc]{JS78, Sell81, Sell85, SA86, ES95, SV97, DS00, Sell03}
The full source code is available, and may be downloaded and installed
as described in \S\ref{sec.download}.  The code may be copied freely
(see \S1.2), but people publishing results that use the software it
provides should acknowledge the source and cite this paper.

An on-line manual for use of the code is maintained at {\tt
  http://www.physics.rutgers.edu/$\sim$sellwood/manual.pdf}.  Since
this is comprehensive, I give only an outline here.

The quality of any collisionless $N$-body simulation is almost
entirely limited by the number of particles employed, which places a
high premium on efficiency.  For many applications of interest, the 3D
methods described here are tens to hundreds of times faster than the
popular tree codes, and the 2D methods have a still greater
performance advantage.  However, speed comes at the cost of reduced
flexibility and increased danger of compromised results from
inappropriate use.  While the code is neither as versatile nor as easy
to use as other publicly available packages, the careful user should
find the increased quality of the results obtainable is ample
compensation for the effort required to learn how to use it.

\bigskip
\subsection{Contents of the package}
\label{sec.packages}
To make use of an $N$-body code, one must first set up the model to be
evolved and afterwards analyze the behavior of the simulation in order
to draw conclusions.  The \routine{GALAXY} package therefore supplies
three integrated pieces of code:

\begin{enumerate}[A]
\item The most obvious, and in many ways the easiest part to develop
  and describe, is the code to compute the evolution of an $N$-body
  system.  The program \routine{galaxy} advances a pre-created
  file of particles in time for as long as is desired.  This is the
  core of the code, where efficiency is key.  At the end, and at
  intermediate times if desired, it creates a file with the full
  information about the state of the system, from which the evolution
  can be continued.

\item Analysis of the results is both more difficult to develop and to
  describe, largely because almost every project has a different
  scientific purpose, and therefore requires some piece of analysis
  that could be unique.  However, most projects also make use of
  standard analyses of the system state and the code provides many
  such options.  Unlike most other publicly available codes, these
  standard analyses in the \routine{GALAXY} package are computed
  ``on-the-fly'', and the results at each analysis step are saved in a
  file for later examination.  The huge advantage of analysis
  on-the-fly is that the results are much more compact and can
  therefore be saved a frequent intervals without the inconvenience of
  large data storage and time-consuming analysis programs.  The
  principal disadvantage is, of course, that a simulation must be
  rerun if a quantity was not measured in the first run.  However, the
  analysis options are now well enough developed that this happens but
  rarely.

\item By far the most difficult part of a successful $N$-body project
  is the creation of the desired initial set of particles from which
  to start the simulation.  Assigning initial positions and velocities
  of a specified system that is to be in equilibrium is generally
  hard, and it becomes especially difficult for a system, such as a
  disk galaxy, with multiple mass components: a disk, bulge, and a
  halo.  The code package includes sophisticated software that can be
  used in some cases, but naturally not every possible option is
  anticipated.
\end{enumerate}

\bigskip
\subsection{License}
\label{sec.license}
\routine{GALAXY} is a package of free software, distributed under the
GNU General Public License.\footnote{see
  http://www.gnu.org/copyleft/gpl.html} This implies that the software
may be freely copied and distributed.  It may also be modified as
desired, and the modified versions distributed as long as any changes
made to the original code are indicated prominently and the original
copyright and no-warranty notices are left intact.  Please read the
General Public License$^2$ for more details.  Note that the author
retains the copyright on the code.

\bigskip
\subsection{Source code}
\label{sec.code}
The code is written in standard Fortran (f90) and should compile
without difficulty.  The basic \routine{galaxy} program uses FFTs from
the open access package {\tt FFTPAK} \citep{Schw82} that are bundled
with the source code.  The post-run analysis software requires the
open access package {\tt PGPLOT}.\footnote{ available from
  http://www.astro.caltech.edu/$\sim$tjp/pgplot/}

However, some parts of the code do make use of the NAG
library,\footnote{see
  http://www.nag.co.uk/numeric/fl/FLdescription.asp} which is a
commercial package.  Many institutes and universities may have it and
even a very old version would be adequate.  The package attempts to
manage this aspect in as user-friendly manner as possible through the
use of an environment variable.

The initialization program \routine{begin} and the main $N$-body
\routine{galaxy} program should execute without the NAG library for
most applications.  {\bf Thus most $N$-body simulations can be run
  without the NAG library}.  Other goodies, such as parts of the
analysis code, and essentially all the set-up software, make more
extensive use of the NAG library and without it, those wishing to use
these powerful tools will have to make a significant programming
effort to substitute routines from no-cost libraries.

The main programs \routine{begin}, \routine{start}, \routine{galaxy},
and \routine{finish} do not make use of MPI calls and can therefore be
used only for running on a single processor.  There is a separate set
of executables \routine{begin\underscore mpi},
\routine{start\underscore mpi}, \routine{galaxy\underscore mpi}, and
\routine{finish\underscore mpi} that enable calculations on multiple
processors, which therefore require a system with some version of MPI.
Even though these executables must be linked with MPI, they can also
be used for running on a single processor if desired.  Throughout the
documentation, the names of these programs are used without the
extension \routine{\underscore mpi} to imply both versions.

\bigskip
\subsection{Downloading and compiling the code}
\label{sec.download}
The installation assumes {\tt PGPLOT}$^3$ and a version of {\tt MPI}
are installed.  The NAG$^4$ library is optional but, without it, many
powerful features will be unavailable.

Proceed with the following steps:

\begin{enumerate}
\item To download the software, go to the website \hfil\break
  \hfil\break \hbox{\hspace{2cm} {\tt
      http://www.physics.rutgers.edu/galaxy}} \hfil\break \hfil\break
  Enter your email address, click {\tt download} and the file {\tt
    galaxy.tgz} will download to your system automatically.  (The
  email address will enable you to be notified when updates are
  available.)

\item This gzipped $\sim6$MB tar file can be moved from your Download
  directory to some convenient location in your directory tree and
  then unpacked with the command:

\noindent{\hspace{2cm} {\tt tar xzf galaxy.tgz}}

and all the files of the package will be created in the directory
\routine{pckg} and subdirectories thereof.  The scripts and Makefile
assume that the user will be using \routine{gfortran}, and will need
to be edited if a different compiler is desired.

\item Build libraries from the source code.  (These two {\tt rebuild}
  scripts use {\tt gfortran} by default.)

\noindent{\hspace{2cm} {\tt cd pckg/lib14}}

\noindent{\hspace{2cm} {\tt ./rebuild}}

\noindent{\hspace{2cm} {\tt cd ../sfftpak}}

\noindent{\hspace{2cm} {\tt ./rebuild}}

These scripts should create the two object libraries
\routine{main14.a} and \routine{sfftpak.a} in the main directory
\routine{pckg}.

\item To run in parallel, the package will need an implementation of
  MPI.  The scripts and Makefile assume {\tt
    OpenMPI}\footnote{available from
    http://www.open-mpi.org/software/ompi/v1.8/} and they will need to
  be edited if a different installation of MPI is to be used.  The mpi
  compiler is assumed to be {\tt /opt/openmpi/gnu/bin/mpif90}

\noindent{\hspace{2cm} {\tt cd ../lib14\underscore mpi}}

\noindent{\hspace{2cm} {\tt ./rebuild}}

This script should create the additional object library
\routine{main14\underscore mpi.a} in the directory \routine{pckg}.

\noindent{\hspace{2cm} {\tt cd ../progs}}

\item To inform the compiler of the location of the PGPLOT object
  library, edit the line that begins {\tt pgplot\underscore lib =} in
  the Makefile (in the {\tt progs} subdirectory).

\item Decide whether the NAG library is available.  Skip to step 7 if
  NAG is unavailable, otherwise there are two additional steps:

\begin{enumerate}[a]
\item Edit the line that begins {\tt nag\underscore lib =} in the
  Makefile (in the {\tt progs} subdirectory) to inform the compiler of
  the location of the NAG object library.

\item Set the environment variable {\tt NAG} with the following command:

\noindent{\hspace{2cm} {\tt setenv NAG nag}}

If this environment variable is not set, or if it is set to any other
value, then a restricted fraction of the code will be compiled.  While
these programs are generally expected to execute normally, they could
terminate gracefully with an explanatory message if a NAG routine is
required for some particular application.

\end{enumerate}

\item Compile the code:

\noindent{\hspace{2cm} {\tt make}}

\noindent{\hspace{2cm} {\tt make clean}}

The {\tt make clean} will not only tidy up the directory, but will move
all the executables to the user's directory \routine{\$HOME/bin},
which is presumed to be in the user's {\tt PATH}.  To update your path,
issue the command:

\noindent{\hspace{2cm} {\tt rehash}}

\end{enumerate}

Depending upon whether NAG is available, the following executables of
the \routine{GALAXY} package should now be ready for use:

\noindent{\tt analys, begin, begin\underscore mpi, cold, compress, corrplt,
  dfiter, dflook, finish, finish\underscore mpi, galaxy, \hfil\break
  galaxy\underscore mpi, isotropy, merge, modefit, plotpft, ptest,
  ptest\underscore mpi, smooth, start, start\underscore mpi, weed}

Without NAG, only the following restricted set of programs can be used:

\noindent{\tt analys, begin, begin\underscore mpi, finish, finish\underscore
  mpi, galaxy, galaxy\underscore mpi, merge, ptest, ptest\underscore mpi}

Executables without the {\tt \underscore mpi} suffix are for single
processors only.  The use of these programs is described in the
on-line manual {\tt
  http://www.physics.rutgers.edu/$\sim$sellwood/manual.pdf}.

\begin{figure}[t]
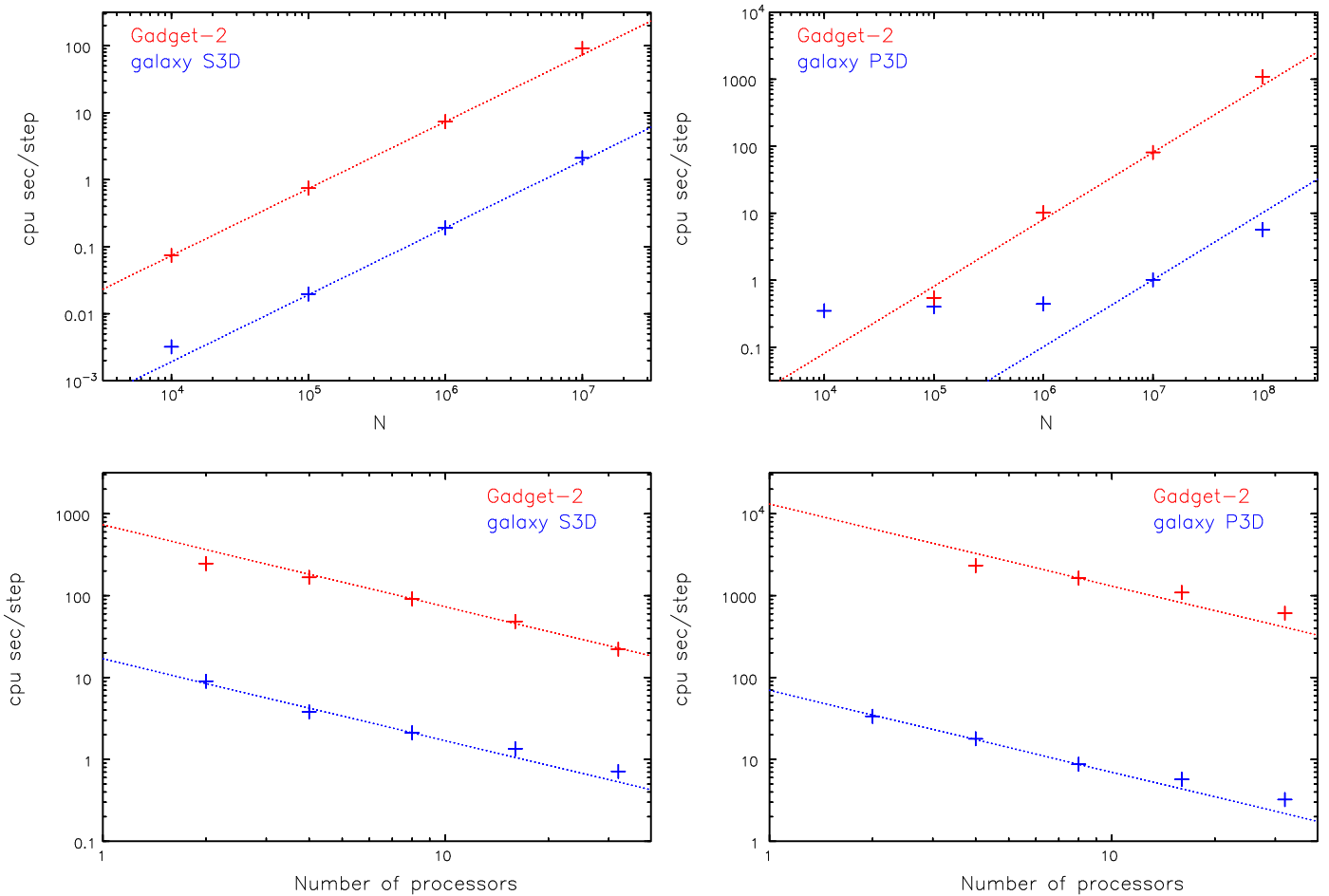

\begin{center}
\hbox to \hsize{\hfil
\includegraphics[width=.49\hsize,angle=0]{tim-s3d.ps} \hfil 
\includegraphics[width=.49\hsize,angle=0]{tim-p3d.ps}}
\end{center}
\caption{CPU timings for \routine{galaxy} and \routine{Gadget-2}.  The
  left panels are for a spherical model, the right panels for a disk.
  The upper left panel shows the scaling with particle number $N$ for
  8 cores while the upper right uses 16 cores.  The lower left panel
  shows scaling with the number of processors for $N=10^7$, while the
  right is for $N=10^8$.  The dotted lines indicate slopes $=\pm1$,
  and are not fits to the data.}
\label{fig.benchmarks}
\end{figure}

\bigskip
\subsection{Benchmark tests}
\label{sec.benchmarks}
\citet{SD06} reported that \routine{galaxy} ran $\sim 37$ times faster
than the non-public tree code \routine{PKDGRAV} \citep{Stad01} on a
single processor.  Here I report further benchmark tests against the
public tree code \routine{Gadget-2} \citep{Spri05}.  While
\routine{Gadget-2} has been used quite extensively to simulate
isolated galaxies, it was designed for cosmic structure formation
calculations that cannot be simulated with \routine{galaxy}.  This
comparison is simply to highlight the relative efficiency of
\routine{galaxy} for isolated galaxies, and is in no way intended as a
criticism of \routine{Gadget-2}.

Fig.~\ref{fig.benchmarks} shows the measured cpu (wallclock) time for
identical files of equal-mass star particles in simulations with
\routine{galaxy} and \routine{Gadget-2}.  The timings shown are cpu
seconds per step averaged over 100 time steps.  The upper panels show
scaling with $N$ for 8 cores (left) and 16 cores (right), while the
lower panels show scaling with the number of cores for $N=10^7$ (left)
and $N=10^8$ (right).  All other numerical parameters were held fixed
as the particle number and number of cores were varied.  The opening
angle used in \routine{Gadget-2} was 0.5 radians and the particle
softening length was $0.05a$.

The left-hand panels are for an equilibrium Plummer sphere, of core
radius $a$, using the spherical grid option of \routine{galaxy} with
surface harmonics $0 \leq l \leq 4$ included in the force
determination.  There were 4 time-step zones, and the grid was
recentered every 8 time steps.  The right-hand panels are for a
thickened KT disk run using a cylindrical polar grid of size $N_R
\times N_a \times N_z = 86 \times 128 \times 125$, with sectoral
harmonics $0 \leq m \leq 8$ included in the force determination.  In
this case, there were 3 time-step zones, and the grid was not
recentered.

As may be seen from the Figure, \routine{galaxy} runs $\sim 50$ times
faster than \routine{Gadget-2} for spherical models consistently for
all $N$ and regardless of the number of processors employed.  For the
disk model, the speed advantage of \routine{galaxy} over
\routine{Gadget-2} gradually improves as $N$ is increased.  Because
there is a fixed overhead for the grid force determination, the
running time for \routine{galaxy} barely increases until $N \ga 10^6$,
and for very large $N$, \routine{galaxy} can be $\ga 200$ times faster
than \routine{Gadget-2}.

\bigskip
\subsection{Disclaimers}
\label{sec.disclaim}
This free software comes with ABSOLUTELY NO WARRANTY.  Users are
welcome to make use of it, change it, and to redistribute it under
certain conditions -- see {\tt http://www.gnu.org/licenses/} for
details.

It is not expected that a would-be user could download the software
provided and quickly run his/her desired simulation using it.  The
basic \routine{galaxy} program should prove relatively easy to use as
a ``black box'' and I provide a few test cases whose results the user
should be able to reproduce without additional coding.  But many
possible pitfalls could vitiate results from hasty use in a new
application and I anticipate that a capable prospective user would to
need to gain many weeks of experience using the package before feeling
sufficiently confident that results are publishable.

While the author welcomes reports of bugs in the distributed software,
which he will attempt to fix, he is unable to provide help to modify
it for some other application.  The user should run the standard tests
to ensure that the package has been properly installed, and report to
the author if any of these tests fail on the user's system.  But he
cannot offer help beyond that point.

\section*{Acknowledgments}
The author wishes to thank all the many collaborators and graduate
students with whom he has worked over the years, many of whom have
made helpful suggestions to improve or add to the software.
Especially noteworthy contributions have been made by my thesis
advisor Richard James, whose Poisson solver \citep{Jame77} is bundled
verbatim with my code, and my students Neil Raha, David Earn, Victor
Debattista, and Juntai Shen.  The SCF option in the code originates
from \citet{HO92} and I am grateful for their permission to include
some of their code.  Joel Berrier has assisted by testing pre-release
versions of the package.  This work was also supported in part by NSF
grant AST/1108977.



\addcontentsline{toc}{section}{References}
\begin{thebibliography}{}

\def\aap{A\&A}
\def\aj{AJ}
\def\an{Astron.\ Nach.}
\def\apj{ApJ}
\def\apjl{ApJL}
\def\apjs{ApJS}
\def\apss{Ap.\ Sp.\ Sci.}
\def\araa{ARAA}
\def\fcp{Fund.\ Cosmic Phys.}
\def\jcop{J. Comp.\ Phys.}
\def\mnras{MNRAS}
\def\newa{New. Astron.}
\def\pasj{PASJ}
\def\PhD{PhD.\ thesis}
\def\phr{Phys.\ Rep.}
\def\nat{Nature}
\def\raa{RAA}
\def\rpp{Rep.\ Prog.\ Phys.}
\def\sci{Science}

\smallskip
\bibitem[\protect\citeauthoryear{Debattista \& Sellwood}{2000}]{DS00}
Debattista, V. P. \& Sellwood, J. A. 2000, \apj, {\bf 543}, 704

\bibitem[\protect\citeauthoryear{Earn \& Sellwood}{1995}]{ES95}
Earn, D. J. D. \& Sellwood, J. A. 1995, \apj, {\bf 451}, 533

\bibitem[\protect\citeauthoryear{Hernquist \& Ostriker}{1992}]{HO92}
Hernquist, L. \& Ostriker, J. P. 1992, \apj, {\bf 386}, 375

\bibitem[\protect\citeauthoryear{James}{1977}]{Jame77}
James, R. A. 1977, \jcop, {\bf 25}, 71

\bibitem[\protect\citeauthoryear{James \& Sellwood}{1978}]{JS78}
James, R. A. \& Sellwood, J. A. 1978, \mnras, {\bf 182}, 331

\bibitem[\protect\citeauthoryear{Schwarztrauber}{1982}]{Schw82}
Swarztrauber, P. N. 1982, in {\it Parallel Computations}, ed.\ G. Rodrigue (Academic Press), p51

\bibitem[\protect\citeauthoryear{Sellwood}{1981}]{Sell81}
Sellwood, J. A. 1981, \aap, {\bf 99}, 362

\bibitem[\protect\citeauthoryear{Sellwood}{1985}]{Sell85}
Sellwood, J. A. 1985, \mnras, {\bf 217}, 12

\bibitem[\protect\citeauthoryear{Sellwood}{2003}]{Sell03}
Sellwood, J. A. 2003, \apj, {\bf 587}, 638

\bibitem[\protect\citeauthoryear{Sellwood \& Athanassoula}{1986}]{SA86}
Sellwood, J. A. \& Athanassoula, E. 1986, \mnras, {\bf 221}, 195

\bibitem[\protect\citeauthoryear{Sellwood \& Debattista}{2006}]{SD06}
Sellwood, J. A. \& Debattista, V. P. 2006, \apj, {\bf 639}, 868

\bibitem[\protect\citeauthoryear{Sellwood \& Valluri}{1997}]{SV97}
Sellwood, J. A. \& Valluri, M. 1997, \mnras, {\bf 287}, 124

\bibitem[\protect\citeauthoryear{Springel}{2005}]{Spri05}
Springel, V. 2005, \mnras, {\bf 364}, 1105

\bibitem[\protect\citeauthoryear{Stadel}{2001}]{Stad01}
Stadel, J. G. 2001, \PhD, University of Washington.

\end{thebibliography}
\end{document}